\documentstyle[pre,aps,tighten,epsf,floats]{revtex}
\begin{document}
\renewcommand
\baselinestretch{2}
\large
\title{\LARGE Motion in a rocked ratchet with spatially periodic friction  \vspace{0.5in}}
\author{\large Debasis Dan,$^{1,*}$ Mangal C. Mahato,$^{2}$ and
A. M. Jayannavar$^{1,\dagger }$ \vspace{0.2in}}
\address{\large $^1$ Institute of Physics, Sachivalaya Marg,
Bhubaneswar 751005, India}
\address{\large $^2$ Department of Physics, Guru Ghasidas University,
 Bilaspur 495009, India}
\maketitle
\begin{abstract}
\vspace{1.0in}
\large

We present a detailed study of  the transport and energetics of a Brownian particle
moving in a periodic potential in the presence of an adiabatic external periodic drive.
The particle is considered to move in a
medium with periodic space dependent friction  with the same periodicity  as
that of  the potential but with a phase lag. We obtain several results, most of them
arising due to the medium being inhomogeneous and are sensitive to the phase lag.
When the potential is symmetric we show that efficiency of energy transduction can be
maximised as a function of noise strength or temperature. 
However, in the case of asymmtertic potential the temperature may or may not facilitate the
energy conversion but current reversals can be obtained as a function of temperature and the
amplitude of the periodic drive. The reentrant behaviour of current can also
be seen as a function of
phase lag.  \\
PACS number(s): 05.40+j, 05.60+w, 82.20Mj

\end{abstract}
\newpage 
\section{Introduction}

The search for the possibility of unidirectional motion in a periodic system without
the application of any obvious bias is of current research 
interest\cite{han,jul,mag,bar,doe,ast,mie,chen,aj,ma,ch,por,Mill,Jay,Par}.
 Such possibility requires 
the system to be out of equilibrium in order for the process to be consistent with the
second law of thermodynamics. Several physical models have been proposed to obtain such
motion. In all the models noise plays the central role. One of the most discussed models is
the one in which an asymmetric periodic potential system is adiabatically rocked\cite{mag,doe}
 by applying
constant forces $F$ and $-F$ at regular intervals of time. One obtains unidirectional motion because
in such a system the current $j(+F)\ne -j(-F)$. Eventhough the time averaged applied force over
a period vanishes the averaged current $<j>=0.5[j(+F)+j(-F)]$ becomes finite in the presence of noise (
thermal fluctuations). Moreover, the average current $<j>$ was found to peak at an intermediate
noise strength (or temperature). In this model it has been further shown that by suitably
choosing the asymmetric periodic potential one may obtain current reversal\cite{Hanggi} as a function of
temperature provided the rocking frequency is high. Similar results, however, can be obtained
in the presence of a unbiased colored noise instead of the oscillating force. There are several other
interesting models to obtain unidirectional motion including models where potential barriers themselves
are allowed to fluctuate\cite{ast} or models wherein symmetric potential system is driven by temporally
asymmetric forces\cite{aj,ma,ch}, etc. 

The result that thermal noise helps to obtain unidirectional current in a periodic system was quite 
important. But later on it was pointed out that obtaining mere current does not necessarily mean
that the system does work efficiently\cite{tak1}. Doing work involves flow of current against 
load and hence one
must, in the spirit of the model, obtain current up against a tilted (but otherwise periodic)
potential system. Analysis shows, however, that the efficiency of an adiabatically rocked system
(ratchet) monotonically decreases with temperature. Therefore though such a ratchet system helps extract
a large amount of work at an intermediate temperature ( where the current peaks) the work is 
accomplished at  a larger 
expense of input energy; thermal fluctuation does not facilitate efficient energy conversion in
this model ratchet system. In a subsequent work\cite{tak2} this deficiency was rectified but in 
a different model
wherein the asymmetric potential oscillates in time, instead of its slope being changed (rocked)
between $+F$ and $-F$ adiabatically. In both these models the friction coefficient was constant and
uniform in space. The present work makes a detailed study of the rocked ratchet system with 
nonuniform friction coefficient which varies periodically in 
space\cite{Mill,Jay,Par,But,Kam,Zhao,Bao,dan1,dan2,Mah,Kam2,takn,Sancho}.
 
In this work we take the friction coefficient to vary with the same periodicity as the 
potential but with a 
phase difference, $\phi$. The phase difference $\phi$, the amplitude $\lambda$ of variation of 
friction coefficient, 
the amplitude $F_0$ of rocking, the load, etc. affect the functioning of the ratchet in an intricate and 
nontrivial manner. The two of the important results we obtain are: (1) The efficiency of the adiabatically
rocked ratchet shows a peak as a function of temperature, though the peak (which may or may not exist
in case of spatially asymmetric potentials) position does not coincide 
with the temperature at which the current peaks, and (2) the current could be made to reverse its
direction as a function of noise strength and the amplitude $F_0$ even at low frequencies of rocking.
These attributes are solely related to the medium being inhomogeneous with space dependent friction.
It is worth noting that the introduction of space dependent friction, though does not affect the
equilibrium properties (such as the relative stability of the locally stable states), changes the 
dynamics of the system 
in a nontrivial fashion.  Recently it has been shown that these systems exhibit noise induced
stability, it shows stochastic resonance in washboard potentials without the application of external
periodic input signal,\cite{dan1,dan2} and also unidirectional motion in periodic symmetric potential 
(non ratchet-like) systems.\cite{Par,Zhao,Bao,dan3}

In the next section we describe our model and obtain an expression for current and efficiency in the
quasi-static limit. In Sec. III we present our results.

\section{Equation of motion in inhomogeneous systems}

The nature of correct Fokker-Planck equation in the presence of space-dependent diffusion 
coefficient (inhomogeneous medium) was much debated earlier. Later on the correct expression 
was found from a microscopic treatment of system-bath coupling. The motion of an overdamped 
particle, in a potential $V(q)$ and subject 
to a space dependent friction coefficient $\gamma (q)$ and an external
force field $F(t)$ at temperature $T$ is described by
the Langevin equation \cite{Par,dan1,dan2,Mah,Sancho}
\begin{equation}
  \label{Langv}    
  \frac{dq}{dt} = - \frac{(V'(q)-F(t))}{\gamma (q)} - k_{B}T
  \frac{\gamma '(q)}{[\gamma (q)]^{2}} + \sqrt{\frac{k_{B}T}{\gamma
      (q)}} \xi (t) ,
\end{equation}
where $\xi (t)$ is a randomly fluctuating Gaussian white noise with zero
mean and correlation :\\  $ <\xi (t) \xi (t')> = 2 \delta (t-t')$. Here $< .. >$ denotes 
an ensemble average over the distribution of the fluctuating noise $\xi(t)$. The
primes in Eq. (1) denote the derivative with respect to the space variable $q$. 
It should be noted that the above equation involves a multiplicative noise with
an additional temperature dependent drift term. The additional term turns out
to be essential in order for the system to approach the correct thermal equilibrium
state. We take
$V(q) = V_{0}(q) + qL$, where $V_{0}(q + 2n\pi) = V_{0}(q) = -Vsin(q)$,
$n$ being any natural integer. $L$ is a constant force ( load)
representing the slope of the washboard potential against which the
work is done. Also, we take the friction coefficient $\gamma (q)$ to
be periodic :\\ $\gamma (q) = \gamma _{0}(1 - \lambda \sin (q + \phi))$, 
where $\phi$ is the phase difference with respect to $V_{0}(q)$. The
equation of motion is equivalently given by the Fokker-Planck equation
\begin{equation}
  \label{Fok_Plk}      
  \frac{\partial P(q,t)}{\partial t} = \frac {\partial}{\partial q} \frac{1}{\gamma
    (q)} [k_{B}T \frac {\partial P(q,t)}{\partial q} + (V'(q) - F(t)
  )P(q,t)] . 
\end{equation}
This equation can be solved for the probability current $j$ when $F(t)
= F_{0}$ = constant, and is given by \cite{dan1,dan2,Ris}
\begin{equation}
  \label{curr}      
  j = \frac{k_{B}T ( 1 - \exp(-2 \pi(F_{0} - L)/k_{B}T))}{\int _{0}^{2
      \pi}\exp(\frac{-V_{0}(y) + (F_{0} - L)y}{k_{B}T}) dy
    \int_{y}^{y+2\pi}\gamma(x)\exp \frac{V_{0}(x) - (F_{0} 
      - L)x)}{k_{B}T}} dx .
\end{equation}
In the presence of space dependent friction and the phase lag 
$\phi \neq 0, \pi$ and in the absence of load 
$j(F_{0}) \neq 
-j(-F_{0})$  even for a spatially periodic symmetric potential. Thus when the 
system is subjected to an external ac field $F(t)$ the unidirectional particle flow 
(or rectification of the current) takes place. The phase lag $\phi$ brings in the
intrinsic asymmetry in the dynamics of the system. 
When an externally applied ac force changes slowly enough (quasi-static or
adiabatic limit) i.e., when the time scale of variation of $F(t)$ is much
larger compared to any other time scales involved in the system we can readily
obtain an expression for the unidirectional current.
For a field $F(t)$ of a square wave
amplitude $F_{0}$, an average current over the period of oscillation
is given by, $<j>\; =\; \frac{1}{2}\,[j(F_{0}) + j(-F_{0})]$. This 
particle current can even flow against the applied load $L$ and
thereby store energy in useful form. In the quasi-static limit
following the method of stochastic energetics\cite{seki} it can be shown \cite{tak1,tak2} 
that the input energy $E_{in}$ ( per unit time) and the work $W$ ( per unit time) 
that the ratchet system extracts from the external noise are given by 
$E_{in} = \frac{1}{2} F_{0}[j(F_{0})-j(-F_{0})]$ and $W =
\frac{1}{2}L[j(F_{0})+j(-F_{0})]$ respectively. Thus the efficiency 
( $\eta$) of the system to transform the external fluctuation 
to useful work is given by 
\begin{equation}
  \label{effi}         
  \eta = \frac{L[j(F_{0}) + j(-F_{0})]}{F_{0}[j(F_{0}) - j(-F_{0})]}.
\end{equation}

Henceforth all our variables like $<j>, E_{in}, W, F_0, T$ are made dimensionless.
The amplitude of potential $V$ is set to unity as all other energy scales are
scaled with respect to $V$. We evaluate $<j>, W, E_{in}$, and $\eta$ numerically\cite{int}
using Eq. (3). 

\section{Results and Discussions}

First, we present our results for average (net) unidirectional current in a symmetric 
periodic potential induced by adiabatic rocking and in the absence of load.
 We emphasize here that to obtain 
these currents the system must be inhomogeneous. The phase lag $\phi$ (except for
$\phi =0,\pi$, for which unidirectional current is not possible) plays an important 
role in determining the direction and magnitude of $<j>$. For $2\pi > \phi > \pi$,  in the
presence of external quasistatic force $F(t)$ and in the absence of load $L$, we have
a forward moving ratchet (current flowing in the positive direction) and for 
$0 < \phi < \pi$ we have the opposite. For instance, if we examine the effect of
friction coefficient close to the minimum of the potential two different situations
are encountered depending on the value of $\phi$. When $\phi > \pi$ and $F_0 > 0$
the particle experiences lower friction near the barriers in the direction of
acquired velocity. The situation is reverse when $F_0 < 0$.
 From Eq. (3) it follows that in a static force 
$F_0$, $j(F_0)\ne
-j(-F_0)$ , hence rectification of current occurs in the presence of external adiabatic drive.
Moreover, it should be emphasized that the magnitude of current or mobility, in the 
static field $F_0$, depends 
sensitively on the potential and the frictional profile over the entire period. 
Depending on the system parameters the current or mobility can be much larger
or smaller than the current or mobility of a particle moving in a homogeneous
medium characterised by the space averaged frictional coefficient. In fact, in
the intermediate values of temperature and $F_0$ the mobility can be made much
larger than their asymptotic limits. This leads to stochastic resonance in
a washboard potential in the absence of ac signal \cite{dan1,dan2}.

In Fig. 1 the average unidirectional current $<j>$ is presented as a function of
$\phi$ and $T$. For this figure the value of amplitude of square-wave ac field
$F_0 = 0.05$, and the amplitude of frictional modulation $\lambda = 0.9$. It 
can be seen that $<j>$ changes sign as a function of $\phi$ and current exhibits 
either a minimum or a maximum as a function of temperature depending on the value
of $\phi$. At the two limits of temperature ($0$ and $\inf$) the currents vanish.
This is the case for the value of $F_{0}$ less than the critical value $F_{c}$ of $F_{0}$
where the barrier to motion in either direction vanishes. In our case the 
critical value of $F_{0}$ is equal to 1. This stochastic resonance-like phenomenon
has been observed in rocked ratchet systems characterised by asymmetric periodic
potentials \cite{mag,doe,Hanggi}. From the contours of the plot it is clear that as we move away from 
phase shift $\phi = \pi$ positive (or negative) direction the temperature at which
maxima (minima) occurs shifts to a larger value and the absolute value of the 
current at the peak decreases. However, the present symmetric potential 
situation does not lead to multiple current reversals as a function of temperature
in the quasi-static limit of an external drive.

The current as a function of $T$ and $F_{0}$ is shown in Fig. 2 for $\phi = 1.3\pi$ 
and $\lambda = 0.9$. For smaller fields $F_0$ compared to the critical field $F_{c}$
the current exhibits a maximum as a function of temperature. As we increase $F_{0}$
the temperature at which the peak occurs decreases. For fields larger than $F_{0}$
the current, however, decreases monotonically with temperature because the barrier to 
motion disappears. In this high field region mobility of a particle decreases with
increase of temperature \cite{dan2}. However, as we increase the field the net current
$<j>$ monotonically changes and saturates to a finite value. This is in 
contrast $<j>$ vanishes for large $F_{0}$ in the  ratchet subjected to a rocking force in the
absence of space dependent friction.  In 
Fig. 3 we have plotted $<j>$ versus $<F_{0}>$ and $\phi$ for fixed values of $T=0.5$
and $\lambda=0.9$. It can again be seen clearly that the current monotonically
varies and saturates to a value given by $-\frac{\lambda}{2}sin(\phi)$ independent
of temperature. This result \cite{dan2} follows from the analysis of Eq.(3). As expected
current reversal can be seen as a function of $\phi$ for large $F_{0}$. It can also be 
verified that 
the current $<j>$ increases monotonically as a function of the amplitude
$\lambda$ of the friction coefficient and hence we do not present variation of
$<j>$, etc., with $\lambda$.

We now discuss the efficiency of a symmetric periodic potential system in the
presence of space dependent friction driven by an adiabatic periodic field.
The efficiency of such a system with uniform friction has been studied
earlier and it has been shown that temperature does not facilitate the
efficiency $(\eta)$ of energy conversion in the system \cite{tak1}. To calculate $\eta$
we make use of Eqs. (3) and (4). In our analysis the load $L$ is applied against 
the direction of net current (in the absence of load). In this situation
particle current can flow against the applied load $L$ less than some
critical value $L_{c}$ thereby storing energy in useful form. For $L>L_{c}$
one cannot talk meaningfully the concept of efficiency as the current
flows in the direction of the load and hence no storage of useful
energy takes place. In Fig. 4 we have plotted efficiency $\eta$, input energy
$ E_{in}$ and work done $W$ (scaled up by a factor 60 for convenience 
of comparison) 
as a function 
of $T$ for the parameter values, $F_{0}=0.5$, $\phi=1.3\pi$, $\lambda=0.9$, 
and the load
$L=0.04$. The figure shows that the efficiency exhibits a maximum as
a function of temperature indicating that thermal fluctuation
facilitates energy conversion. This in contrast to the case of
uniform friction coefficient where $\eta$ decreases monotonically
with the increase of temperature in the same adiabatic limit \cite{tak1}. It is to 
be mentioned that the temperature corresponding to the maximum efficiency
is not the same as the temperature at which the average
current $<j>$ becomes maximum in the absence of load. 
The temperature at which
the extracted work maximizes is not the same as the temperature at which
the efficiency becomes maximum for the same parameter values. The input
energy increases with temperature monotonically and saturates at the high 
temperature limit. $\eta$, $W$, and $E_{in}$
show similar qualitative behaviour for other parameter values.
The above important observation of temperature facilitating the energy
conversion is applicable for the spatially symmetric potential. In
general in adiabatically rocked systems with frictional nonuniformities
the increasing thermal noise need not increase the efficiency. The 
efficiency is sensitive to the qualitative nature of the periodic 
potential (and also to the nonuniformity of friction). For instance, 
asymmetric potential exhibits quite complex
behaviour of $\eta$ and $<j>$. To illustrate this we take 
 $V_{0}(q) = -\sin q -\frac{\mu}{4}\sin
2q$ ( where $\mu$ lies between -1 and 1, and is the asymmetry
parameter). With this potential we discuss three separate cases:
Case A - system in a symmetric potential ($\mu=0$) in an inhomogeneous
medium ($\lambda \ne 0$), Case B - system in an asymmetric potential 
($\mu \ne 0$) in an inhomogeneous
medium ($\lambda \ne 0$), and Case C - system in an asymmetric potential 
($\mu \ne 0$) in a homogeneous medium ($\lambda=0$).

In Fig.5, we have presented results of $\eta$ versus $T$ for all the three 
cases described above. For this we have taken $F=0.5$, $L=0.002$, and 
$\phi = 1.3\pi$. For case A $\lambda=0.9$, for case B $\mu=0.08$ and 
$\lambda=0.9$, and for case C $\mu=0.08$. As discussed earlier for the case A
temperature maximizes the efficiency. Case C, where the medium is 
homogeneous, efficiency monotonically decreases with temperature. These
observations have been emphasized in earlier literature. The case B,
 where potential
asymmetry and frictional inhomogeneity are present, the efficiency
decreases monotonically in this parameter regime. In general whether
the temperature facilitates the energy conversion in case B depends 
sensitively on the system parameters. In some limited parameter range
the peaking behaviour is seen as in Fig.6. The parameter 
values are indicated in the caption. The presence of asymmetry in the 
potential may or may not help in enhancing the efficiency of the 
system. This can be seen from Figs. 5 and 6. In all these cases the 
work done $W$ and the input energy $E_{in}$ show similar qualitative
features as shown in the inset of Fig. 1. Now, we discuss the variation 
of efficiency as a function of load.

On general grounds it is expected that the efficiency too exhibits 
maximum as a function of load. It is obvious that the efficiency
is zero when load is zero. At the critical value $L_c$ (beyond
which current flows in the direction of the load) the value of 
current is zero and hence the efficiency vanishes again.
In between these two extreme values of load the efficiency exhibits
maximum. Beyond $L=L_{c}$ the current flows down the load and therefore
the idea of efficiency becomes invalid. In Fig. 7, we have plotted
$\eta$ versus load for all the three cases for chosen values of
parameters as mentioned in the figure. In all these cases current 
monotonically decreases as a function of load. The work done against 
load $W$ exhibits a maximum as a function of load. The load at which
$W$ shows maximum does not coincide with the load at which $\eta$
becomes maximum. The input energy $E_{in}$ as a function of load varies
non monotonically exhibiting a minimum. However, depending on
the case under consideration the value of the load at which the
minimum in the input energies observed may be larger than $L_{c}$
above which efficiency is not defined. 

In Fig. 8, we have plotted the efficiency versus the amplitude of 
the adiabatic forcing $F_{0}$ for all the three cases. It can be seen
from the figure that for the system in an inhomogeneous medium, namely
for cases A and B, $\eta$ exhibits a maxima and saturate to the same 
value in the large amplitude limit. In contrast, for the case C after
exhibiting maximum $\eta$ goes to zero. This follows from the simple 
fact that in the large amplitude limit in the absence of frictional
inhomogeneities the net unidirectional current tends to zero. The
peculiar feature of saturation of efficiency in  inhomogeneous
media is related to the fact that the average current saturates to a
constant value in the high amplitude limit as discussed earlier. This
somewhat counter-intuitive result is typical to inhomogeneous
media.  Having discussed efficiency of energy conversion we now
study the nature of net current $<j>$ in the presence of spatially
asymmetric potential to examine if current reversals take place
in the adiabatic limit in the absence of load. It is known from the 
earlier literature \cite{Hanggi} that in an adiabatically rocked asymmetric
potential ratchet system net current does not exhibit reversals as
a function of $T$.
In these systems current reversals are possible when the frequency
of the applied ac field is large. We show here that in the presence
of frictional inhomogeneities in addition to asymmetry in the 
potential one can observe current reversal as a function of thermal
noise. In Fig. 9, we have plotted the magnitude of net current $<j>$ 
versus $T$ for all the three cases A, B, and C. The corresponding
parameter values are mentioned in the caption of the figure. The
cases A and C do not exhibit current reversal. This is a general
result independent of parameter values. However, in case B current
reversal is observed. To obtain current reversal both asymmetry in
potential and nonuniform friction coefficient are essential, that is,
current reversals arise due to the combined effect of $\phi$ and $\mu$.
Moreover, it should be noted that to observe current reversals the 
parameter range should be such that the net current in case A is 
in the opposite direction to that in case C. For the case B for which 
current reversal is observed, the plot of efficiency separates into 
two disjoint branches as the load should be reversed keeping the
magnitude same when the current reversal takes place. In the presence
of $\mu \ne 0$ and $\lambda \ne 0$, where the current reversals are 
observed, the efficiency as a function of temperature is less than the
maximum value of efficiency in either of the two cases A and C. That is,
$\eta(\mu \ne 0, \lambda \ne 0) < max[ \eta(\mu=0,\lambda \ne 0),
\eta(\mu \ne 0, \lambda=0)]$. To further analyze the nature of current
reversals, in Fig. 10, we have plotted $<j>$ as a function of $\phi$ and
$T$ for fixed values of $\mu=1$, $\lambda=0.9$, and $F_0=0.5$. From the
contour plots it is clear that as a function of $\phi$ the current
reverses sign twice in the intermediate temperature range. Thus the 
current exhibits reentrant behaviour as a function of phase $\phi$
which is special to the case B. As we decrease the asymmetry in the potential 
the $<j>=0$ contour line shifts towards $T=0$ thereby enhancing the
domain of current reversal to a lower value of temperature as a 
function of $\phi$. As a function of temperature the current reversals
occur in a definite range of phase $\phi$ which , in turn, depends
on other material parameters. The qualitative behaviour of $<j>$
remains unaltered for different $F_{0}$ as long as $F_{0}$ is less than
the critical value. 

In Fig. 11, $<j>$ is plotted as a function of $T$ and $F_0$ for $\mu=1$,
$\phi =1.3\pi$, and $\lambda=0.9$. As opposed to the case of symmetric
potential (Fig. 2), currents in the small temperature regime do exhibit maxima
and then saturate to a constant value as noted earlier. As a function of
$T$ the current exhibits similar features as in the case A (Fig. 2).
Figure 12, shows $<j>$ as a function of $F_0$ and $\phi$ for $T=0.1$, 
$\mu=1$, and $\lambda=0.9$. As opposed to case A (see Fig. 3) $<j>$ shows
current reversal as a function of $F_0$ in the range $0< \phi < \pi$.
However, in the asymptotic limit of $F_0$ current saturates to a value
$-\frac{\lambda}{2}sin(\phi)$ independent of the value of the asymmetry
parameter $\mu$. From the contour plot it follows that as a function of 
phase $\phi$  we observe the reentrant behaviour of current at high 
values of $F_0$. As we decrease $\mu$ the $<j>=0$ contour shifts towards
smaller values of $F_{0}$ thus making it possible to observe the double 
reversals at even smaller values of $F_{0}$. This reentrant behaviour
as a function of $\phi$ and the current reversal as a function of $F_{0}$
is very specific to the case B alone (compare Fig. 3). 

Thus we conclude from our studies that the dynamics of a particle in an
inhomogeneous medium is rich and complex. In the presence of adiabatic
forcing and asymmetry in the potential current reversals can be
observed as a function of $T$ and $F_{0}$. And depending on the system 
parameters thermal fluctuations facilitate the energy conversion. The
above behaviour cannot be seen in the homogeneous medium in the same
adiabatic limit. However, it is possible to observe these  
 in homogeneous
media in the presence of finite frequency ac drive (nonadiabatic regime).
This seems to suggest that $\phi$ may play the characteristic role of
frequency in our model in the absence of nonadiabatic ac drive. This
has been noted earlier in the context of observation of stochastic
resonance phenomena in inhomogeneous media in the presence of static
tilt alone \cite{dan2}. As a function of phase $\phi$, we observe reentrant
behaviour for the current which arises because of interplay between
asymmetry, inhomogeneity, thermal noise, and strength of the adiabatic 
forcing. Some of the phenomena can be understood at best at a qualitative 
level only. The effect of nonadiabatic forcing may be of further
interest. Work on this line is under investigation.

\section*{Acknowledgements} 
            MCM acknowledges partial financial support and hospitality 
from the Institute of Physics, Bhubaneswar. MCM and AMJ acknowledge
partial financial support from the Board of Research in Nuclear
Sciences, DAE, India.            

\hspace{1.8in} \Large REFERENCES \hfill
  
\newpage
\hspace{1.5in} \large  FIGURE CAPTIONS .\vspace{0.5in}
\large

Fig. 1. The net current $<j>$ as a function of $\phi$ 
and $T$, for parameter values  $F_{0} = 0.5$, $\lambda=0.9$. In the
base plane contour of  surface plot $<j>(\phi,T)$ are given,
dotted line indicates $<j> = 0$ contour.

\vspace{.3cm}
Fig. 2. $<j>$ as a function of temperature $T$ and the amplitude of
the rocking force $F_{0}$, for $\phi=1.3\pi$, $\lambda=0.9$.

\vspace{.3cm}
Fig. 3. $<j>$ as a function of  the amplitude of
the rocking force $F_{0}$, and phase $\phi$, for $T=0.5$.In the
base plane contour of  surface plot is given. 

\vspace{.3cm}
Fig. 4. Efficiency $\eta , E_{in}$ and $W$ as a function of $T$ for 
$\phi=1.3\pi$, $F_{0} = 0.5$, $\lambda = 0.9$, and $L=0.4$. $W$ has
been scaled up by a factor $60$ to make it comparable with $\eta$ and
$E_{in}$. Y-axis is in dimensionless units.

\vspace{.3cm}
Fig. 5. Efficiency versus temperature for (i) case A ($\mu =0$, and
$\lambda=0.9$), (ii) case B ($\mu = 0.08$ , $\lambda = 0.9$), and (iii)
case C ($\mu = 0.08$ , $\lambda = 0.0$), for fixed $F_{0}=0.5$, $\phi =
1.3\pi$, and $L=0.002$.

\vspace{.3cm}
Fig. 6. Efficiency $\eta$ versus temperature $T$, for different values
of $\mu$ and for $F_0=0.5$, $\lambda=0.9$, $L=0.1$, and $\phi=1.3\pi$.

\vspace{.3cm}
Fig. 7. Efficiency versus load $L$ for (i) case A ($\mu =0$, and
$\lambda=0.9$), (ii) case B ($\mu = 1.0$ , $\lambda = 0.9$), and (iii)
case C ($\mu = 1.0$ , $\lambda = 0.0$), for fixed $F_{0}=0.5$, $\phi =
1.3\pi$, and $T=0.1$.

\vspace{.3cm}
Fig. 8. Efficiency versus $F_{0}$ for (i) case A ($\mu =0$, and
$\lambda=0.9$), (ii) case B ($\mu = 1.0$ , $\lambda = 0.9$), and (iii)
case C ($\mu = 1.0$ , $\lambda = 0.0$), for fixed $L=0.02$, $\phi =
1.3\pi$, and $T=0.1$.

\vspace{.3cm}
Fig. 9. Current $<j>$ versus temperature  $T$ for (i) case A ($\mu =0$, and
$\lambda=0.9$), (ii) case B ($\mu = 1.0$ , $\lambda = 0.9$), and (iii)
case C ($\mu = 1.0$ , $\lambda = 0.0$), for fixed $F_{0}=0.5$, $\phi =
0.3\pi$, and $L=0.0$.

\vspace{.3cm}
Fig. 10. The net current $<j>$ as a function of $\phi$
and $T$, for parameter values  $F_{0} = 0.5$, $\lambda=0.9$, and $\mu=1.0$.
 In the
base plane contour of  surface plot $<j>(\phi,T)$ are given,
dotted line indicates $<j> = 0$ contour.

\vspace{.3cm}
Fig. 11. $<j>$ as a function of temperature $T$ and the amplitude of
the rocking force $F_{0}$, for $\phi=1.3\pi$, $\lambda=0.9$, and $\mu=1.0$.

\vspace{.3cm}
Fig. 12. $<j>$ as a function of  the amplitude of
the rocking force $F_{0}$, and phase $\phi$, for $T=0.5$, and $\mu=1.0$.
 In the
base plane contour of surface plot $<j>(\phi,T)$ are given,
dotted line indicates $<j> = 0$ contour.

\begin{figure}
  \begin{flushleft}
  \centerline{\epsfysize=25cm \epsfbox{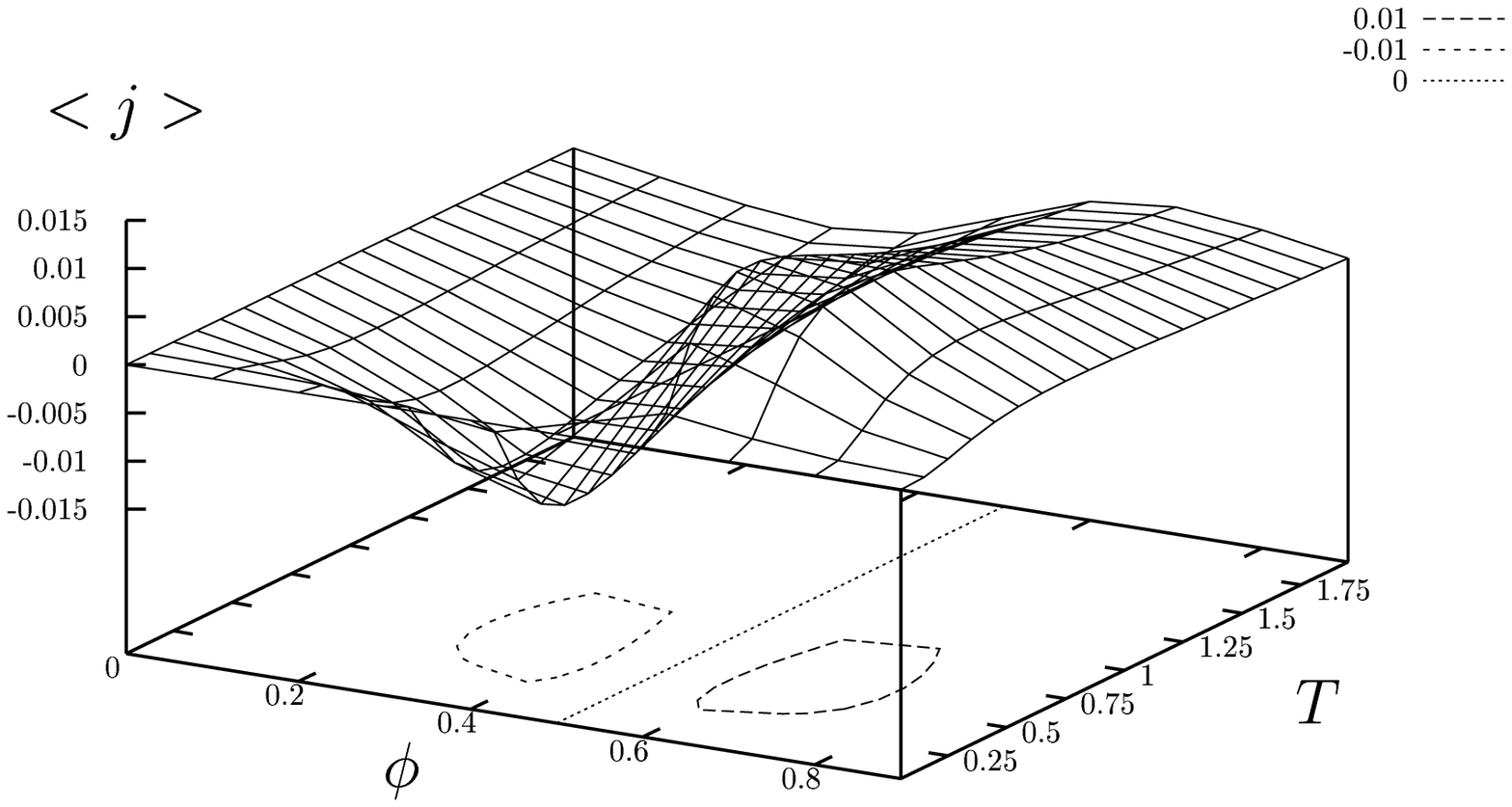}}
  \end{flushleft}
  \vspace{-2.0in}
  \caption{}
\end{figure}

\begin{figure}
  \begin{flushleft}
  \centerline{\epsfysize=25cm \epsfbox{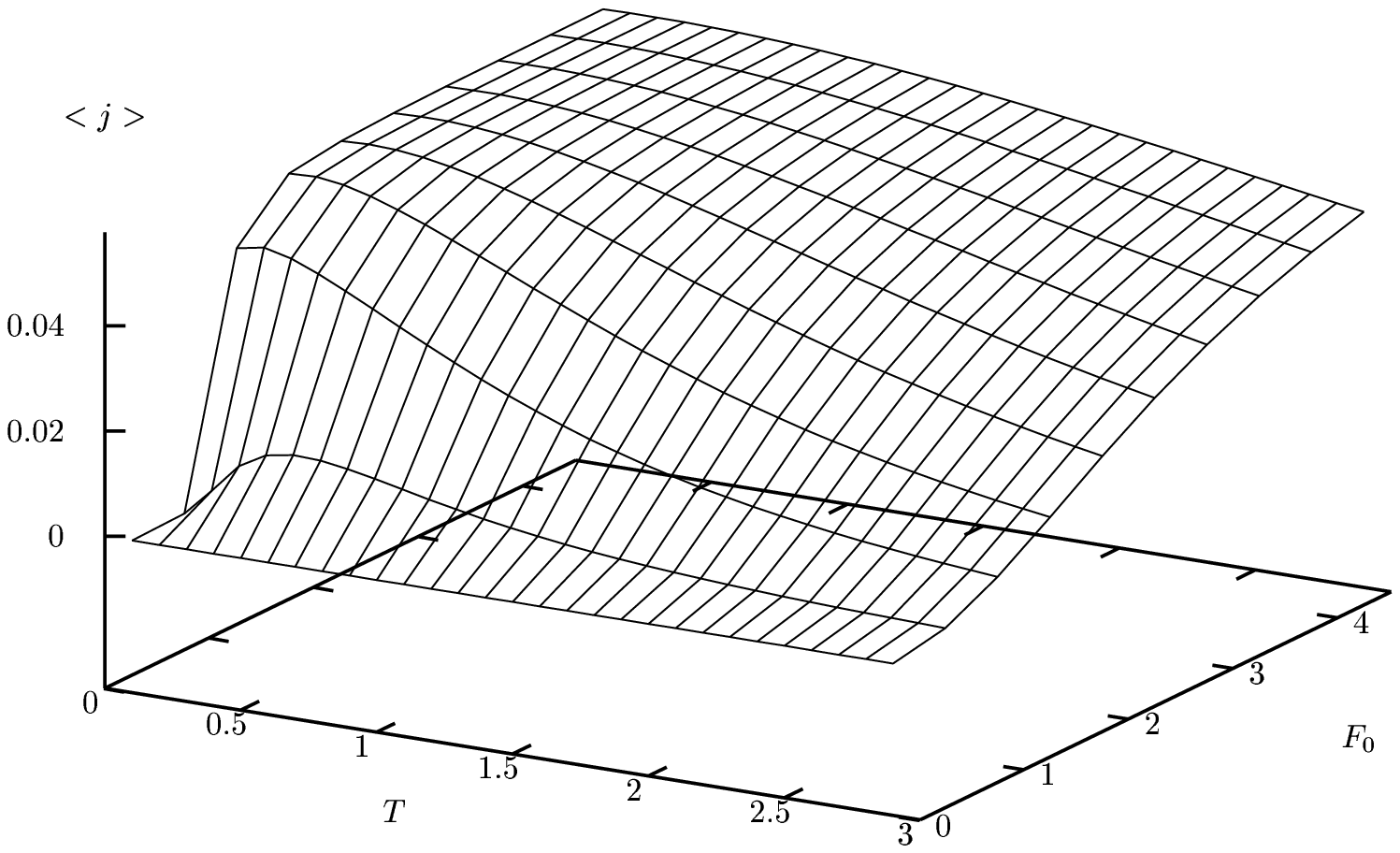}}
  \end{flushleft}
  \vspace{-2.0in}
    \caption{}
\end{figure}

\begin{figure}
  \begin{flushleft}
  \centerline{\epsfysize=25cm \epsfbox{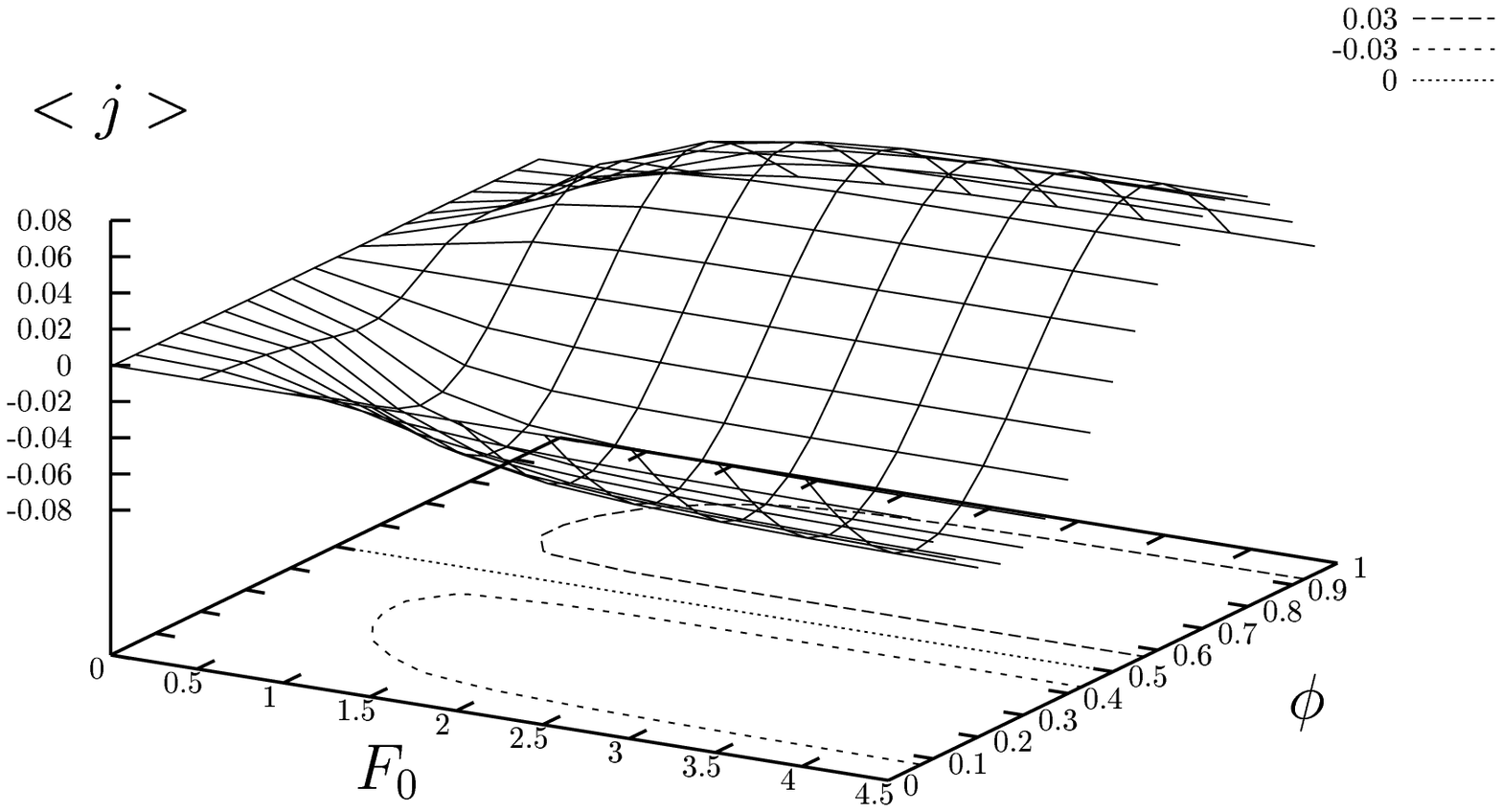}}
  \end{flushleft}
  \vspace{-2.0in}
    \caption{}
\end{figure}

\begin{figure}
  \centerline{\epsfysize=25cm \epsfbox{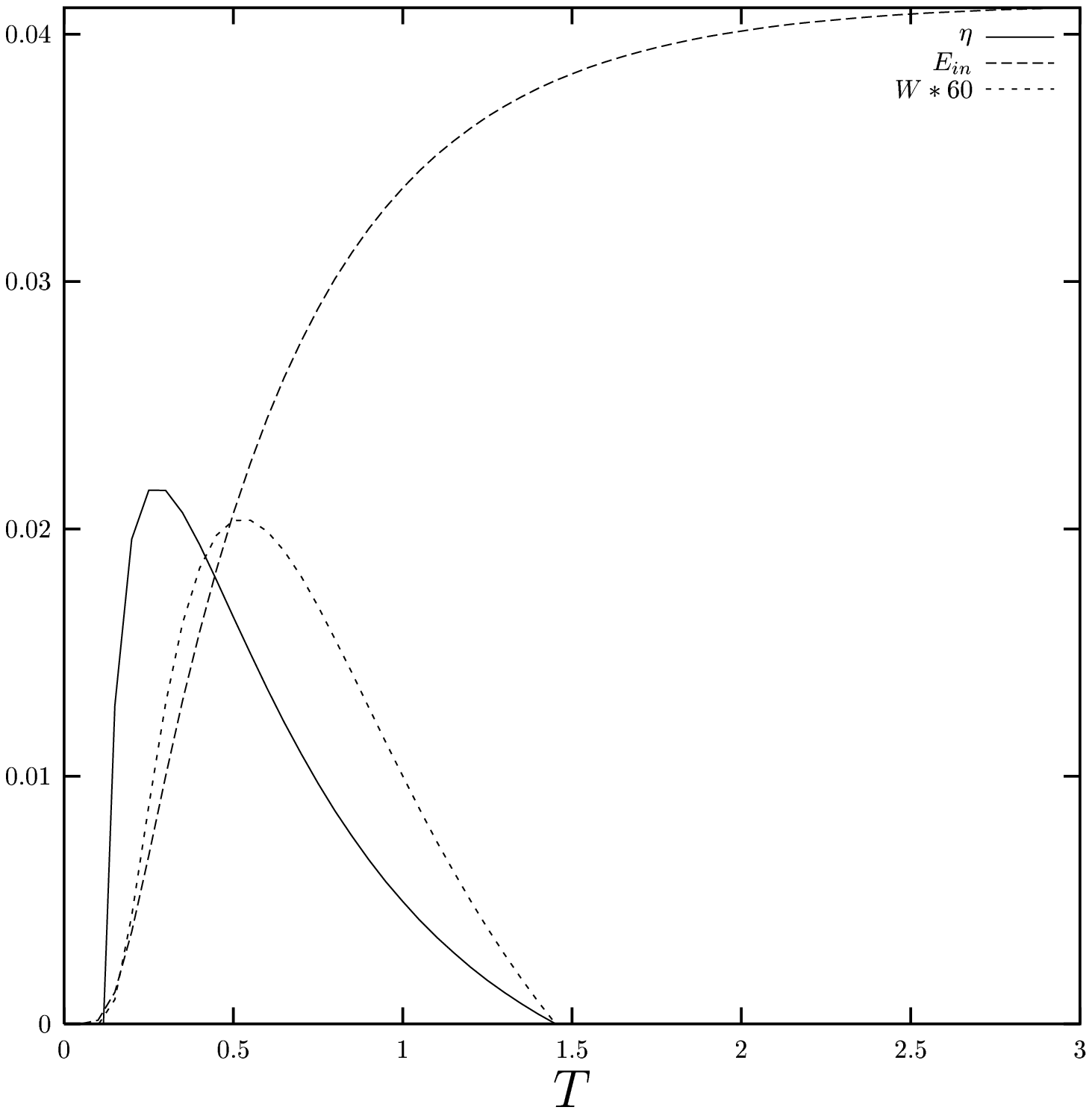}}
  \vspace{-2.0in}
  \caption{}{}{}
  \label{curr_rev}
\end{figure}

\begin{figure}
  \centerline{\epsfysize=25cm \epsfbox{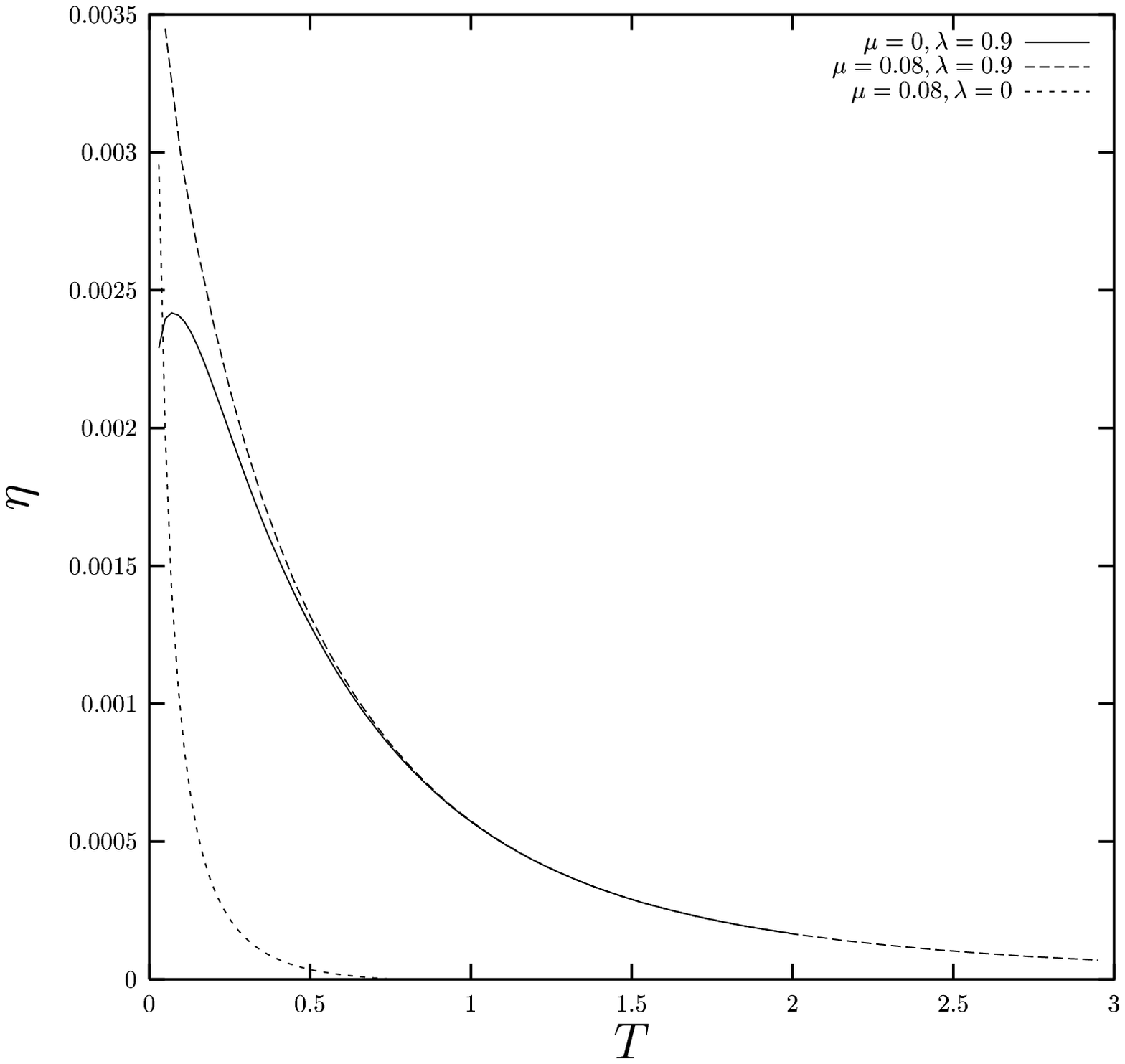}}
  \vspace{-2.0in}
  \caption{}{}{}  \label{curr_rev}
\end{figure}

\begin{figure}
  \centerline{\epsfysize=25cm \epsfbox{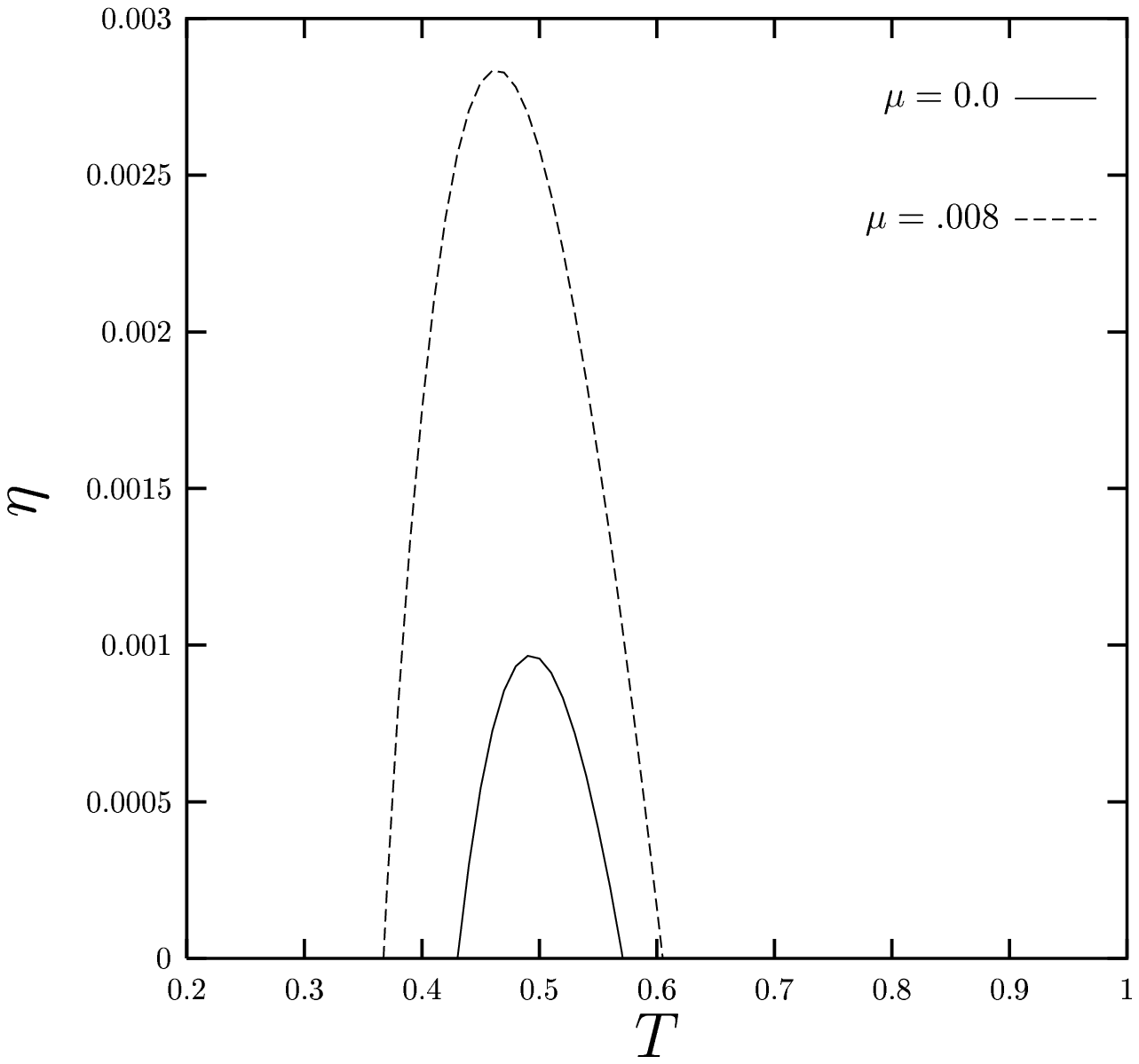}}
  \vspace{-2.0in}
  \caption{}{}{}  \label{curr_rev}
\end{figure}

\begin{figure}
  \center{\epsfysize=25cm \epsfbox{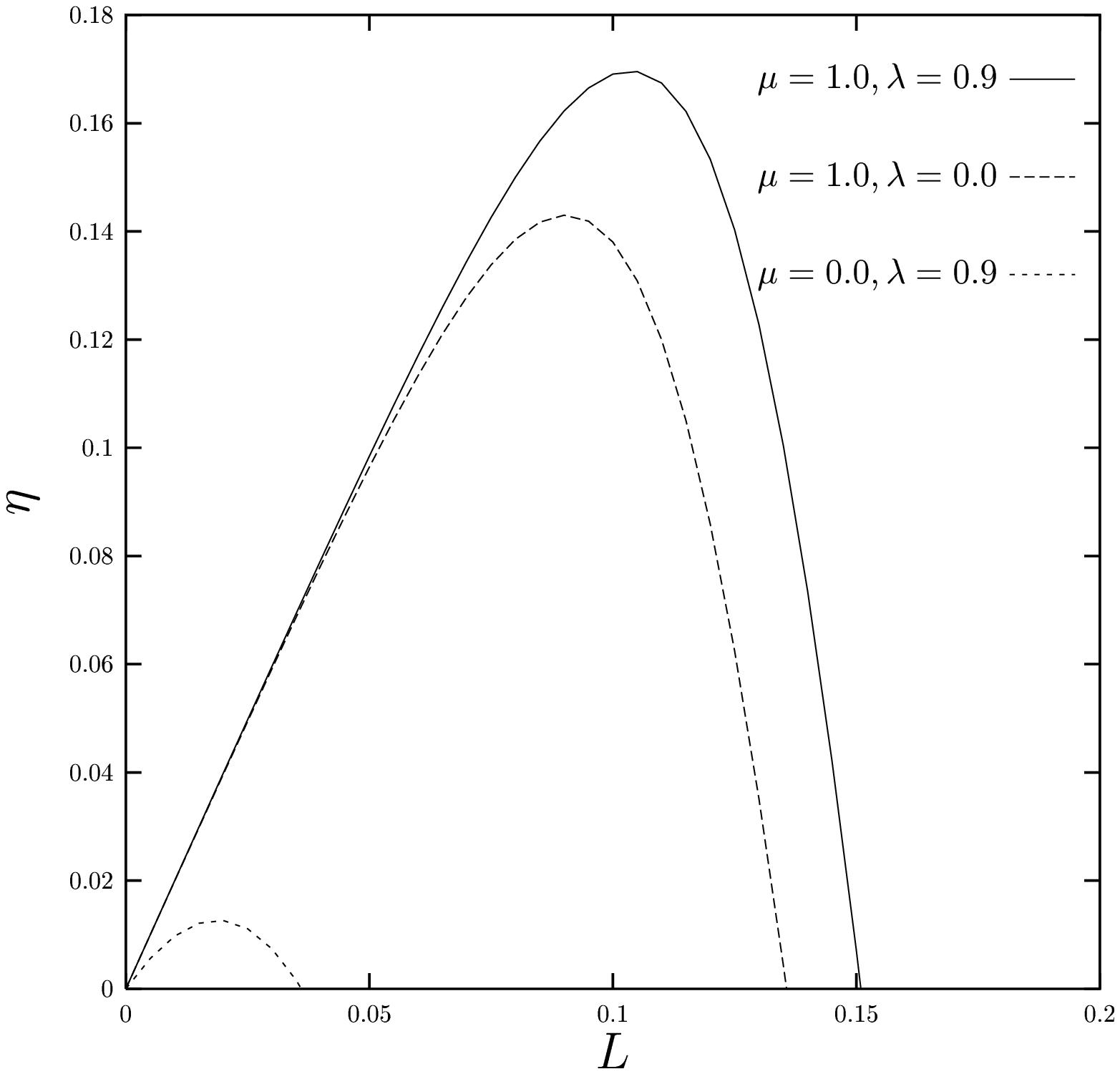}}
  \vspace{-2.0in}
  \caption{}  \label{curr_rev}
\end{figure}

\begin{figure}
  \centerline{\epsfysize=25cm \epsfbox{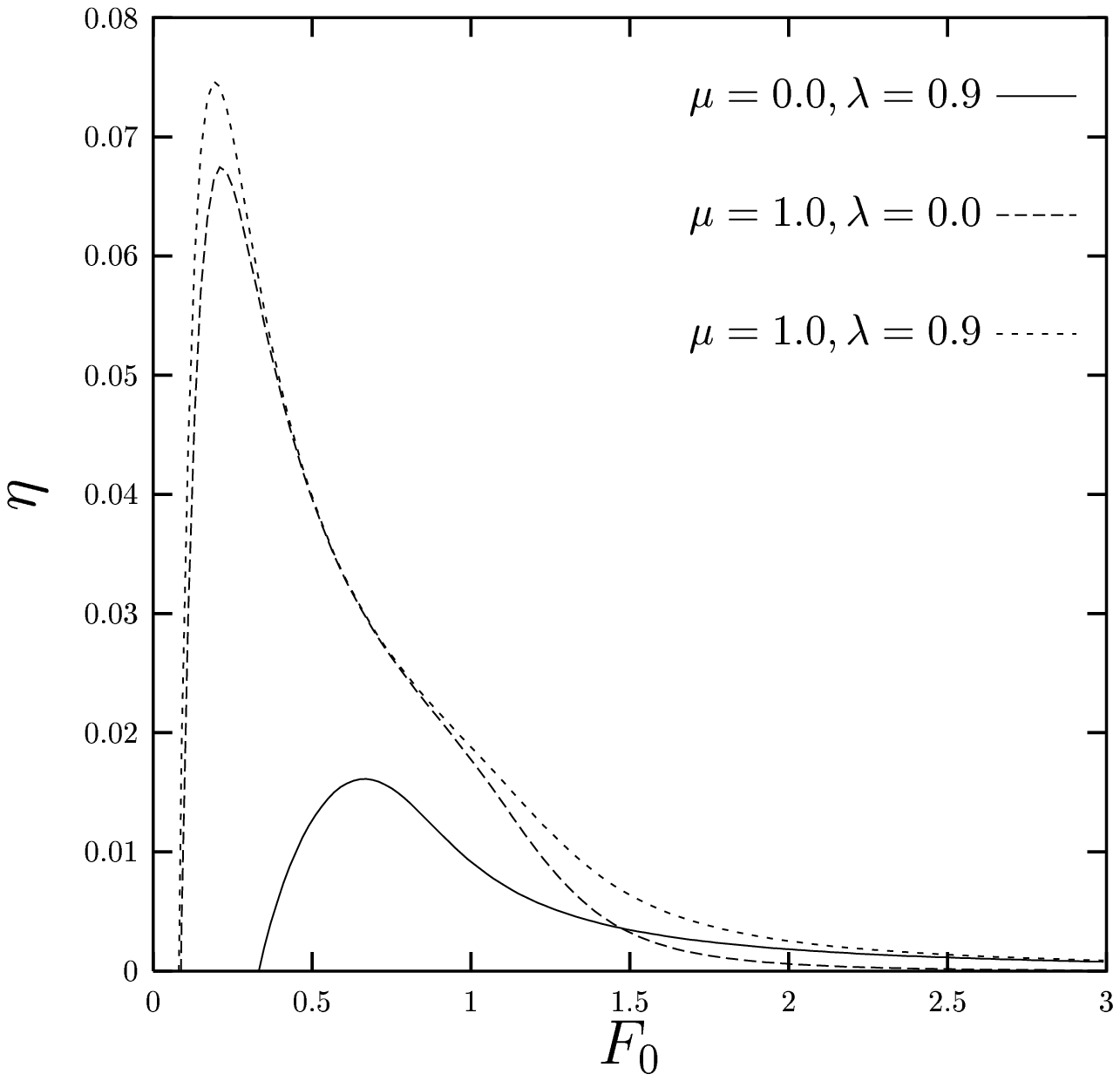}}
  \vspace{-2.0in}
  \caption{}{}{}  \label{curr_rev}
\end{figure}

\begin{figure}
  \centerline{\epsfysize=25cm \epsfbox{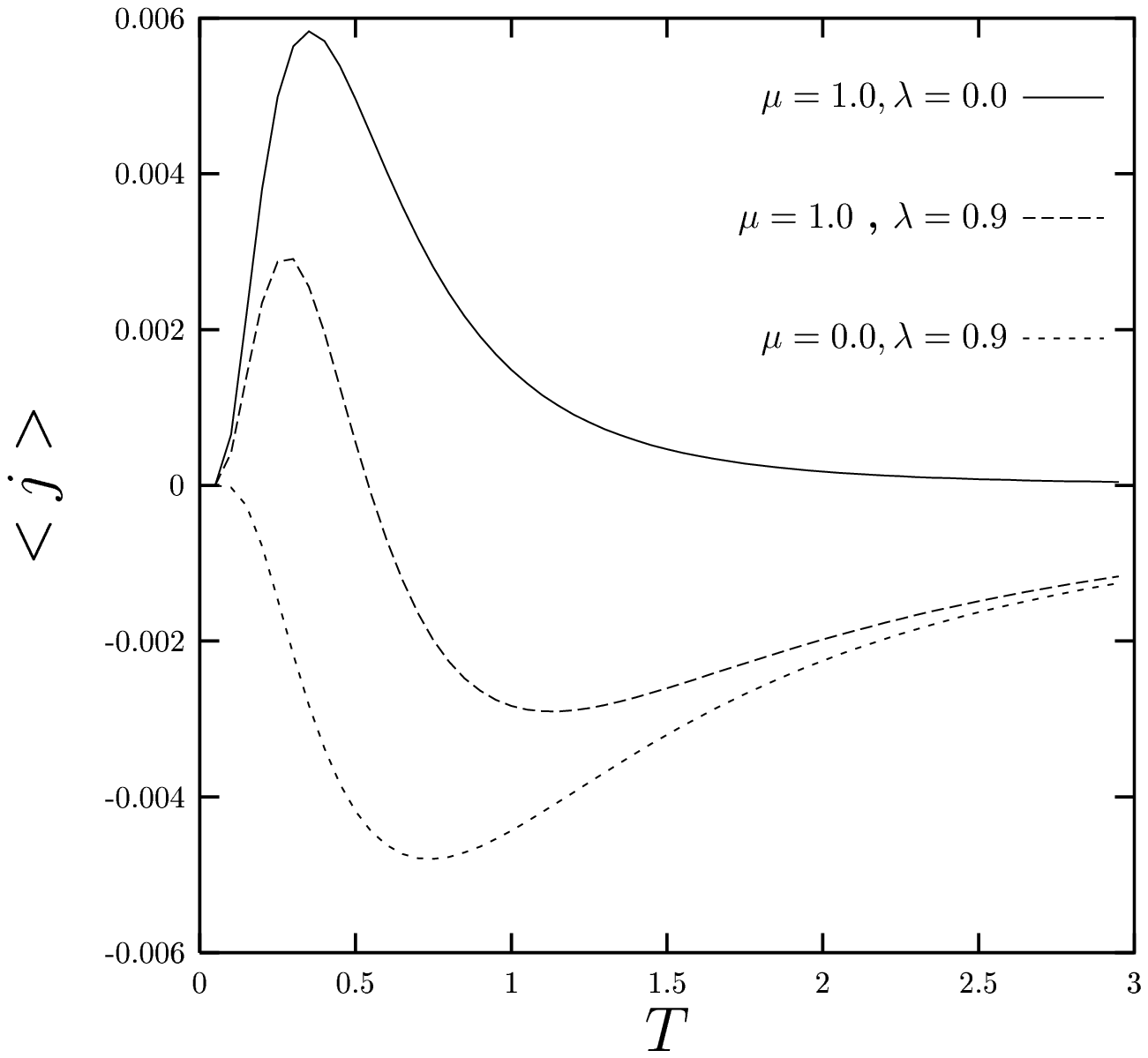}}
  \vspace{-2.0in}
  \caption{}{}{}  \label{curr_rev}
\end{figure}

\begin{figure}
  \centerline{\epsfysize=25cm \epsfbox{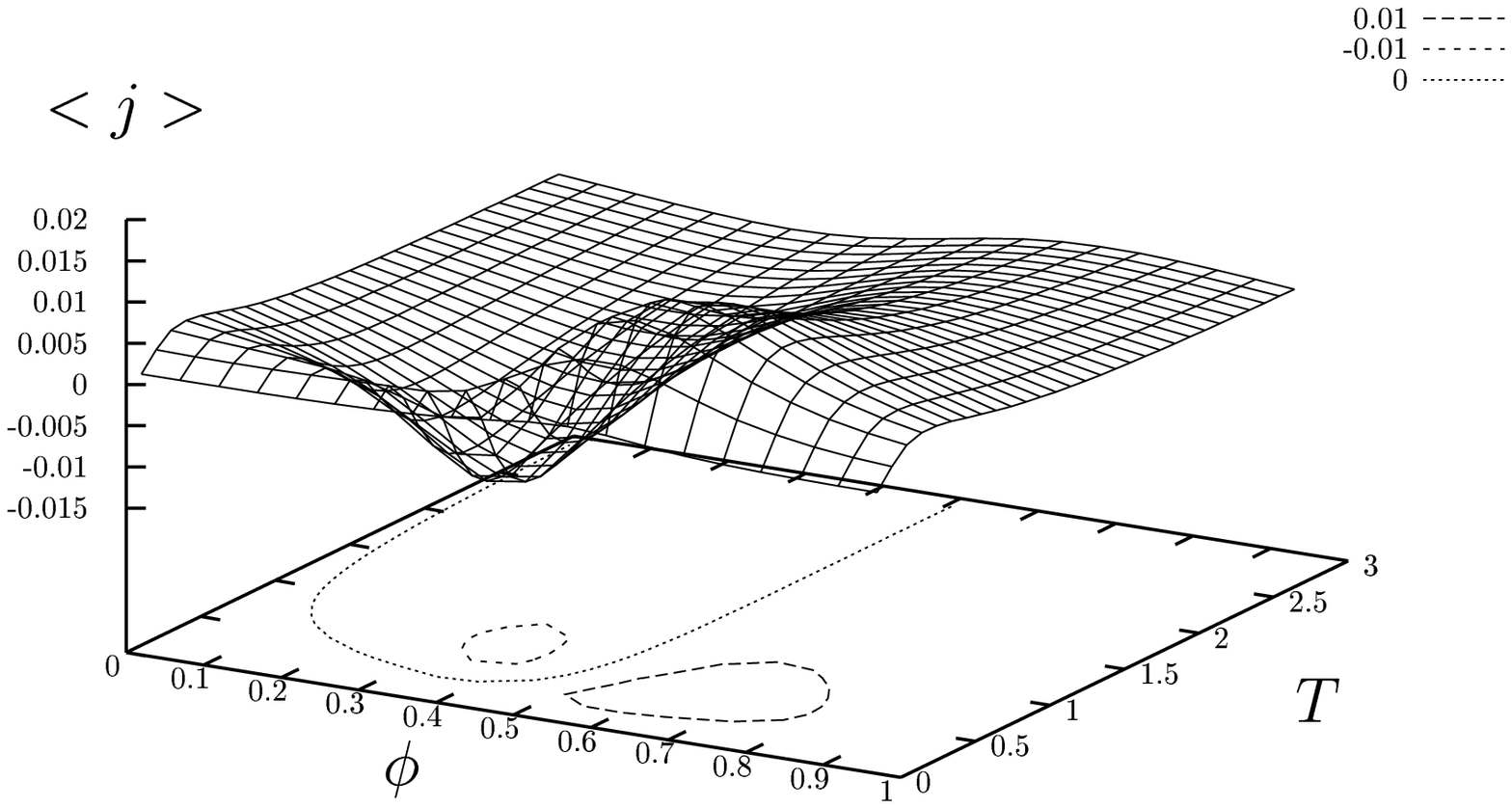}}
  \vspace{-2.0in}
    \caption{}
\end{figure}

\begin{figure}
  \centerline{\epsfysize=25cm \epsfbox{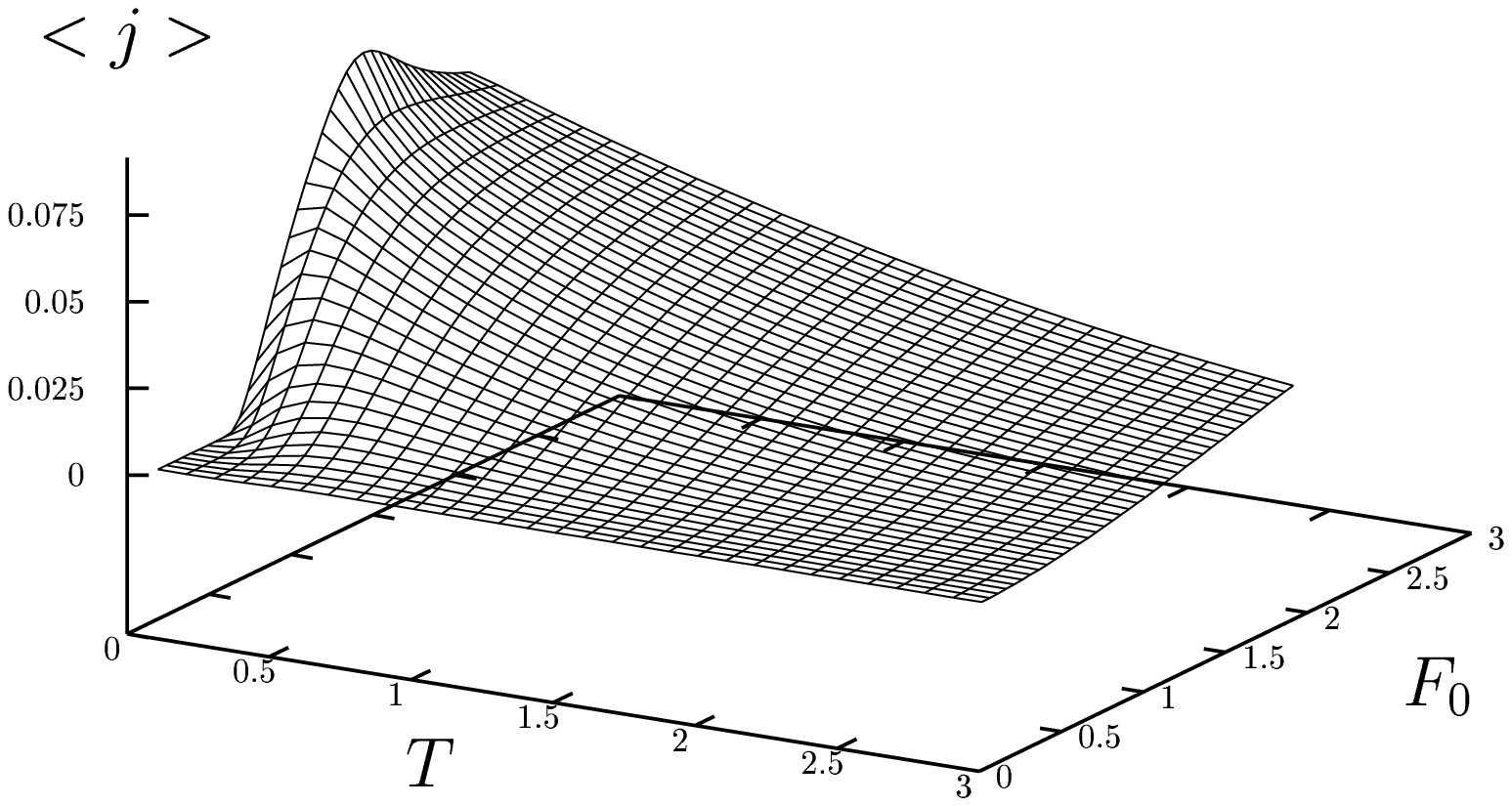}}
  \vspace{-2.0in}
    \caption{}
\end{figure}

\begin{figure}
  \centerline{\epsfysize=25cm \epsfbox{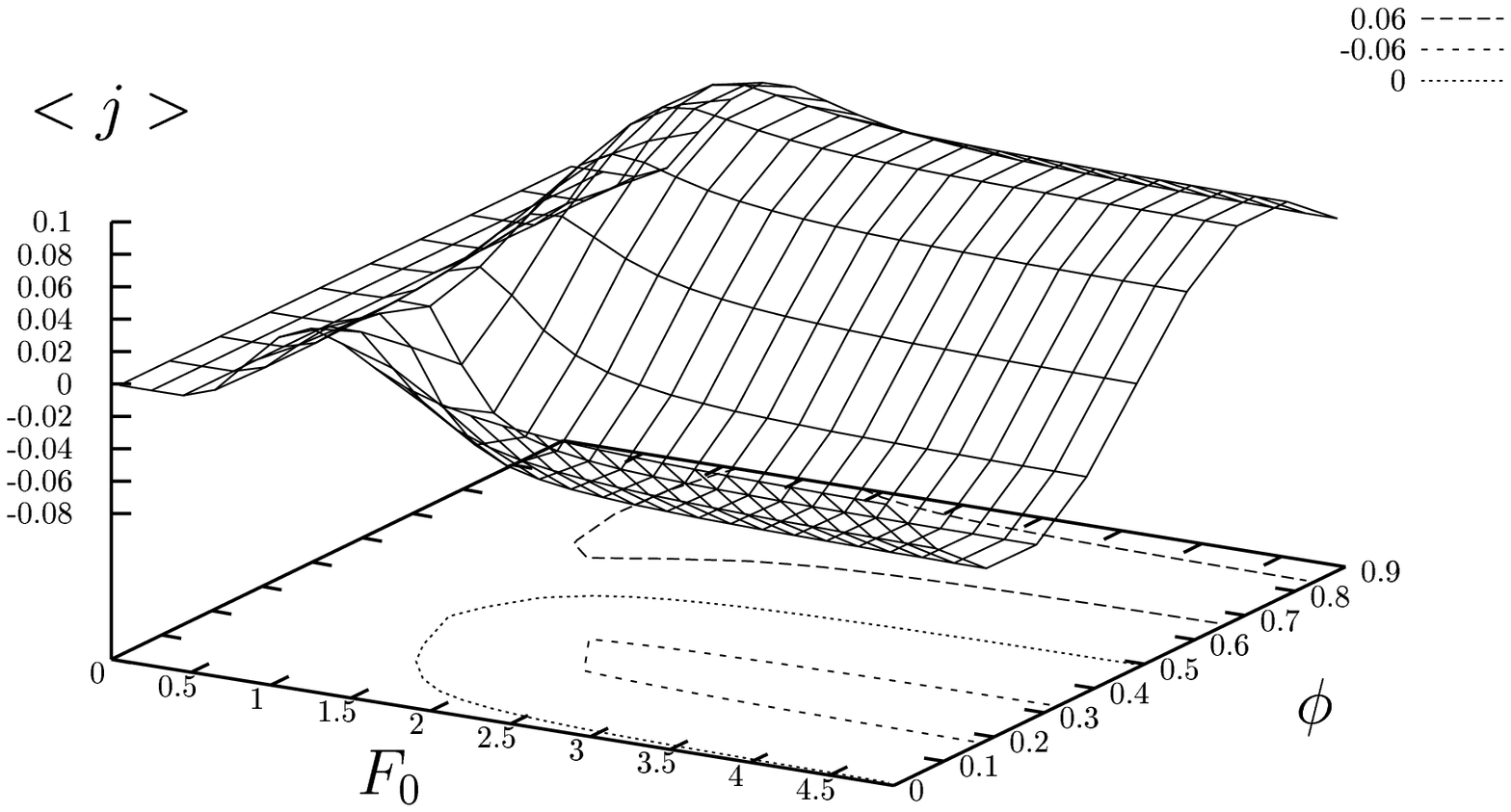}}
  \vspace{-2.0in}
    \caption{}
\end{figure}

\end{document}